\documentclass[useAMS, usenatbib, amssymb]{mn2e}
\usepackage{psfig}
\usepackage{graphicx}
\usepackage{epsf}
\usepackage{bm}
\usepackage{lscape}
\title [Formation and Evolution of W UMa Stars]
{Formation and Evolution of W UMa Stars: Fallacies and Corrections}

\author[Eker et al.]
       {Z. Eker,$^{1, 2} \thanks{E-mail: eker@comu.edu.tr}$
        O. Demircan$^{2}$, S. Bilir$^{3}$
\\ 
  $^1$T\"UB\. ITAK National Observatory, Akdeniz University Campus, 07058 
Antalya, Turkey\\
  $^2$\c Canakkale Onsekiz Mart University, Faculty of Sciences and Arts, 
Ulup\i nar Astrophysical Observatory, 17100 \c Canakkale, Turkey\\
  $^3$Istanbul University, Science Faculty, Department of Astronomy and Space 
Sciences, 34119 Istanbul, Turkey\\
}      
\date{Accepted 2007 month day.
      Received year month day;
      }
\pagerange{\pageref{firstpage}--\pageref{lastpage}}
\pubyear{2007}

\begin{document}

\maketitle

\label{firstpage}
 
\begin{abstract}
The period histograms of eclipsing binaries generated with ASAS data cannot
only be interpreted by orbital evolution. The eclipse probabilities, selection 
effects and space distributions in the solar neighborhood should be considered 
before any interpretations are made. Depending upon physical dimensions 
(total mass and period) of the progenitor stars and the efficiency  
of angular momentum loss (AML) mechanism, a newly formed W UMa type binary 
can be at any age up to several Gyr, and evolution in the contact stage is 
controlled not only by angular momentum and mass loss but also by mass transfer 
between the component stars. Thus, mean life of contact stages should be about 
1.6 Gyr. A different time scale would cause inconsistencies.
\end{abstract}


\maketitle  

\section{Introduction}
Low-mass contact binaries, popularly known as W Ursa Majoris (W UMa) stars, 
are eclipsing binary stars with equally deep eclipses. Observational data 
and theory of W UMa-type contact binaries (WCB) were revised extensively by 
\citet{Moch81, Vilhu81} and \citet{Rucinski82}. According to \citet{Rucinski86} 
the most promising mechanism to form WCB involves orbital angular momentum 
loss (AML) and the resulting orbital decay of detached but synchronized close 
binaries. AML by magnetic braking \citep{Schatzman59,Kraft67,M68} became 
especially popular after Skumanich's (1972) study, which presented observational 
evidence of decaying rotation rates for single stars. Magnetic braking 
and tidal locking were considered as main route forming WCB from the 
systems initially detached but comparable periods \citep{H66, OS70, Vant79, 
VR80, M84, GB88, Mvant91, Stepien95, D99}. However, a small group of very 
young WCBs were found by \citet{Bilir05}. Such very young ($<$ 0.5 Gyr old) 
WCBs were probably formed right at the beginning of the main-sequence or 
during pre-main-sequence contraction phase \citep{Eker06}. 

Debates on the formation mechanisms continue. Referring to the period histograms 
of eclipsing contact, eclipsing semi-detached and eclipsing detached systems of 
the All Sky Automated Survey (ASAS) data, \citet{Paczynski06} have stated that 
``at this time the contact systems seem to appear out of nowhere'' because the 
number of eclipsing detached systems appear insufficient to produce the observed 
number of eclipsing contact systems. On the contrary, the same period 
histograms of ASAS data, and the kinematical ages of W UMa sub groups, which 
were given by \citet{Bilir05}, have been interpreted by \citet{Li07} that they 
claim after a pre-contact duration of 3.23 Gyr, WCBs must be formed from the 
detached progenitors with orbital periods mostly less than 2.24 days and the 
duration of the contact stage is 5.68 Gyr. However, \citet{Bilir05} has shown 
that both very young (age $<$ 1 Gyr) and old W UMa stars coexist.  

The aim of this paper is to show the period histograms of WCBs produced 
from the ASAS data and the kinematical ages of W UMa sub groups formed by 
\citet{Bilir05} according to orbital period ranges are consistent with the 
classical view of most WCBs are formed from detached progenitors of 
comparable periods and mean duration of the contact stage is about 1.6 Gyr. 
Other scenarios with different lifetimes would be inconsistent and/or 
fallacious.

\section{Discussions}
\subsection{Interpretation of the period histograms of ASAS data}

The All Sky Automated Survey (ASAS) is a long-term project, which lasted in three 
phases of operation dedicated to detecting and monitoring of bright stars 
($V \le$ 14$^{m}$) \citep{Paczynski06}. Using a single instrument with an 
aperture of 7 cm, a focal length 20 cm, a standard $V$ band filter and a 
2K $\times$ 2K CCD camera, in phase III among the 50099 variable stars 
distinguished, 11076 eclipsing binaries were identified and period 
histograms of 5384 contact (EC), 2949 semi-detached (ESD) and 2743 detached 
eclipsing binaries (ED) were produced. Studying the period histograms of EC, 
ESD and ED binaries with $|b|>30^{\circ}$, \citet{Paczynski06} have concluded 
that there are comparable numbers of contact and semi-detached systems but 
the relative number of detached systems is inconsistently small as if 
observational data does not support formation of W UMa stars from the 
detached systems of comparable orbital periods.

Relying on the same data, \citet{Li07} argue that the ASAS data supports the 
view of a formation from progenitors of orbital periods less than $P=2.24$ 
days, via angular momentum loss (AML) driven by magnetic stellar winds (MSW). 
The peak value of $P=2.24$ days of the period distribution of ED systems 
is now old and invalid. One of their evidences was the estimated tidal 
locking limit of $P=2.4$ days. Moreover, Fig. 1 of \citet{Li07} appears to be 
altered without any explanation because the relative numbers of ED systems 
with respect to EC and ESD are not the same as in Fig. 6 of \citet{Paczynski06}. 
By comparing orbital and rotation periods, \citet{Demircan06} estimated that 
tidal locking limit is not less than about 70 days in the field 
chromospherically active binaries (CABs) and around 10 days in a younger 
group of CABs.  

Furthermore, it is not correct to explain the peak value of the diagram 
via tidal locking. Not only the peaks, but also the shapes of the period 
distributions of EC, ESD and ED systems in ASAS data should be explained via 
combined causes of eclipse probability and selection effects.

\subsection{Interpretation of kinematical age versus mean mass and periods}
Being unaware of serious inconsistencies, \citet{Li07} established a theory of 
W UMa formation just because the period distribution peak ($P=2.24$ days) of 
ASAS data of ED binaries were found  with a value close to the old 
estimate of the tidal locking limit of \citet{Vant88}. Similarly, just 
because a 3.23 Gyr decaying time for a typical ED to form a typical 
ESD binary by the rates of \citet{Demircan06} is close to the 
kinematical age of the youngest sub-group (3.21 Gyr) of \citet{Bilir05},
 \citet{Li07} claimed that W UMa binaries must have been formed after a
 pre--contact stage of 3.23 Gyr maximum. Moreover, just because the age 
difference between the youngest and oldest groups in \citet{Bilir05} is 
5.68 Gyr, \citet{Li07} adopted the 5.68 Gyr as the mean lifetime 
of W UMa stars. Consequently, mean lifetime (5.68 Gyr) and typical 
pre-contact duration (3.23 Gyr) require mean kinematical ages of oldest 
W UMa stars to be 8.91 Gyr. Inconsistency is clear because the kinematical
 data of \citet{Bilir05}, from which Li et al.'s (2007) theory was 
established, is known to produce 5.47 Gyr for the mean kinematical
 age of the field W UMa stars. However, Li et al.'s (2007) theory 
overestimates mean kinematical ages as [8.91 (oldest) - 3.23 (youngest)]/2 
+ 3.23 = 6.07 Gyr. Moreover, such a theory appears to be established on a wrong 
conception that all stars in different age groups have similar ages which is not 
true. It is possible the oldest group may contain a W UMa star just formed at 
an age of 3.23 Gyr, while the youngest group may contain a W UMa star which is 
of about 9 Gyr of age. Finally, Li et al.'s (2007) theory fails to explain very 
young (ages $<$ 0.6 Gyr) W UMa stars which had been discussed by \citet{Bilir05}.   

Adoption of the 5.68 Gyr lifetime for WCBs by \citet{Li07} implies a scenario 
like this: a detached binary, being eligible to form a WCB, must join to the 
youngest group after a 3.21 Gyr of a detached pre-contact phase. Then, being 
a typical WCB, it must continue secular evolution further by losing AM and 
mass but according to the rates of \citet{Li07} while visiting all groups in 
their Table 2 one by one; after the oldest group, it ends as a fast 
rotating single star. The youngest group with longest periods mostly contains 
WCB of spectral types A or early F. Gradually, the dominant spectral types 
changes to K types at the oldest group. This scenario is like to assume 
main-sequence stars evolve from O types to M types by losing heat, mass and 
AM without leaving the main-sequence. This is inconsistent and false as if 
stellar evolution of stars, which start as O type single stars on the 
main-sequence and finally end as a M type.     

Appearance on some diagrams, e.g. H-R, AM-P, etc diagrams, could be misleading. 
The direction of the evolution needs an independent evidence. For the H-R 
diagram, independent evidence comes from internal structure and evolution 
models, which predicts the direction of the evolution is from the 
main-sequence towards the red giants or super-giants region. Demircan et al.'s 
(2006) method provides direction and amount of the dynamical evolution of CAB 
stars independently. But, the same method is not applicable to WCBs 
\citep{Eker07} because pre-contact detached duration varies. As it is predicted 
from the theory of tidal locking and magnetic braking via magnetic stellar 
winds, a pre-contact stage could take any amount of time within the 
main--sequence lifetime \citep{GB88, Vant88, Stepien95} depending upon 
the initial periods and masses of progenitors. This means that, a detached 
binary, if it is eligible, may join any of the field W UMa sub-groups of 
\citet{Bilir05} by losing mass and orbital AM as having an age within the 
range from zero to several Gyr as suggested by the initial conditions of 
the binary orbit and the rate of AML. 

As also noted by \citet{Bilir05} that the field W UMa sub groups are not 
only populated by systems dynamically evolving from a younger sub group 
to an older sub group. Joining any group may occur unexpectedly by a system 
of any age which just become a WCB. Fitting a linear variation of AM, $P$ or 
$M$ according to the mean kinematical ages of WCB sub-groups may provide a 
rate mathematically but would be meaningless physically. Note that, this is 
not the case for CAB stars since there is no pre-CAB problem. 

It is still not known what would be the correct method to deduce dynamical 
evolution for WCBs on a AM-total mass diagram. The method of \citet{Li07} 
would appear to be wrong because masses and periods of W UMa stars are 
not arbitrary as in the case of the detached systems. There could be 
detached systems all having the same orbital period but their systemic 
masses may vary from half a solar mass to tens of solar masses. This is not 
the case in WCBs since mass contained in Roche lobes is limited; changing the 
mass requires the changing of the orbital period and then AM of 
the system will change accordingly. Consequently, using kinematical 
ages to estimate $dJ/dM$ will be useless to indicate true dynamical evolution 
since $J$ (systemic AM) and $M$ are not arbitrary and time dependence 
cancels when computing $dJ/dM$. More importantly, AM evolution of WCBs is 
not only due to AML through magnetic stellar winds, but mass transfer between 
the components also plays a dominant role which is not easy to handle 
but still must be considered in the evolution of WCBs.

\section{Conclusions}
1) It is possible to model the period histogram of binaries from the ASAS data 
theoretically. Such a model would contain mostly the eclipse probabilities and 
observational selection effects together with estimated true number density 
distribution of binaries with different masses and orbital periods. Such a 
modeling would also be useful to estimate the true distribution which could 
be useful in studying the origins of binaries. 

2) WCBs can be formed at any age depending upon the physical conditions of 
their progenitors such as their periods, component masses and efficiency of the 
AM loss mechanism. Therefore, grouping them into various ages does not 
indicate the younger group is the progenitor of the older group stars. 
If there is equilibrium in the population of WCBs, and if they are mostly 
formed from detached CAB systems, 5.47 Gyr of kinematical age of the field 
WCBs \citep{Bilir05}, and 3.86 Gyr kinematical age of CAB systems indicates 
\citep{Karatas04} a mean life time of the contact stage is about 1.61 Gyr as 
in the pool problems. Otherwise a different lifetime would be inconsistent 
with existing kinematical data.   

\section{Acknowledgments}
Thanks are given to the referee Dr. Slavek Rucinski, who made suggestions 
improved the overall quality of discussions.

\end{document}